# Macroscopically entangled light fields: A quantum laser

Byoung S. Ham

Center for Photon Information Processing, School of Electrical Engineering and Computer Science,
Gwangju Institute of Science and Technology
123 Chumdangwagi-ro, Buk-gu, Gwangju 61005, S. Korea
(Submitted on February 09, 2021; bham@gist.ac.kr)

**Abstract:**
A novel method of macroscopically entangled light-pair generation is presented for a quantum laser using randomness-based deterministic phase control of coherent light in a Mach-Zehnder interferometer (MZI). Unlike the particle nature-based quantum correlation in conventional quantum mechanics, the wave nature of photons is applied for collective phase control of coherent fields, resulting in a deterministically controllable nonclassical phenomenon. For the proof of principle, the entanglement between output light fields from an MZI is examined using the Hong-Ou-Mandel-type anticorrelation technique, where the anticorrelation is a direct evidence of the nonclassical features in an interferometric scheme. For the generation of random phase bases between two bipartite input coherent fields, a deterministic control of opposite frequency shifts results in phase sensitive anticorrelation, which is a macroscopic quantum feature.

**Introduction**

Since the seminal paper by Einstein, Polodsky, and Rosen (EPR) in 1935 [1], the so-called spooky action of nonlocal correlation has been intensively studied for the fundamental understating of quantum mechanics [1-16]. For direct evidence of nonclassical features in entangled photon pairs, the Bell inequality violation [2], Franson-type nonlocal correlation [3], and Hong-Ou-Mandel (HOM) anticorrelation [4] have been investigated over the decades in both noninterferometric [5-9] and interferometric schemes [10-16]. In these studies, not only entangled photon sources from spontaneous parametric down conversion (SPDC) processes [17], but also independent light sources from such as quantum dots and sunlight [18] have been used for demonstrating nonclassical features via coincidence measurements. However, all of these studies have focused on the particle nature of photons, even though coherence is the bedrock for entanglement generation. Providing entangled photon pairs is an essential step toward quantum information processing via controlled not gate operations [19], entanglement swapping [20], quantum teleportation [21], and unconditionally secured key distribution [22]. Multiphoton-based bipartite entanglement of a N00N state [23] or a Schrodinger's cat [24] is essential for quantum sensing applications to beat the standard quantum limits. Unfortunately, however, there is no recipe for entangled photon-pair generation. The generation of macroscopic quantum states with large $N > 100$ may not be technically possible with current technologies [25].

Recently, the fundamental physics of quantumness or nonclassicality has been investigated for both HOM dip [26], photonic de Broglie wavelength [27], and Franson-type nonlocal correlation [28] using the wave nature of photons, where the origin of anticorrelation in a HOM dip is rooted in a $\pi/2$ phase shift between the entangled photon pairs [26,29]. The origin of nonlocal correlation has been discovered in the basis randomness for a coupled bipartite system via quantum superposition [30]. Unlike the particle nature of photons limited to coincidence detection, however, the wave nature of photons emphasizes coherence. Here, coherence represents a typical interference such as in Young's double slits. Such coherence has also been demonstrated in an MZI for single photons [31]. Collective phase control of an atomic ensemble has already been demonstrated for quantum interface [32-36]. Likewise, collective phase control of ensemble photons from a laser is a key technique in the present manuscript, resulting in inherent macroscopic quantum manipulation via the orthonormal basis randomness of the coupled system [27-30]. Here, we present a novel theory of macroscopically entangled light-pair generation using the randomness of the phase basis in an MZI. Considering the coherent de Brogle wavelength (CBW) [27], the origin of macroscopically entangled light pairs is the superposition between MZI phase bases [26-30], where randomness is an essential requirement for $g^{(1)}$ coherence [37]. According to the basic quantum physics, the second-order intensity correlation $g^{(2)}$ is closely related with the first-order corrlation $g^{(1)}$ in coherence optics, where $g^{(2)} = g^{(1)} + 1$ [38]. Here, Heisenberg's uncertainty principle does not limit a quantum mechanically coupled system as it does in EPR [1] and Popper [39].

**Results**

Figure 1 shows schematics of the macroscopically entangled light-pair generation in an MZI by providing its phase basis randomness. As is already known, basis randomness is an essential requirement of quantum superposition between bipartite systems such as in Young's double slits and an MZI [37]. Once basis randomness fails, there is no quantum superposition but instead classical superposition [40]. Here, it should be noted that the conventional understanding of classicality for the individuality of coupled photons has been discussed in Bell's inequality theorem [2]. In that sense, coherence optics may or may not belong to classical physics depending on the phase choice, as discussed for anticorrelation [26,29]. Figure 1 is for pure coherence optics, where the first MZI in Fig. 1a is a preparation stage for the random bases between two input fields $E_1$ and $E_2$ by classically controlling the symmetric phases $\zeta$ and $\zeta'$. The original input field $E_0$ in Fig. 1a is for typical laser light, and a single photon case is also included for the present analysis. For the present scope, however, we set $E_0$ as a commercially available laser light for the discussion of macroscopic quantum features. Figure 1b is a phase-controlled light pulse sequence for $E_1$ and $E_2$, where $E_1$ and $E_2$ are designed to be symmetrically detuned by $\pm\Delta$ (blue and red), respectively, in a frequency domain across the center frequency $f_0$ of $E_0$ (green). Here, $I_j$ represents the corresponding intensity of the field $E_j$, where the detuned fields ($E_1$ and $E_2$) are alternatively coupled with the original field $E_0$ (see Figs. 1c and 1d). For example, if $E_1$ ($E_2$) is turned on, $E_2$ ($E_1$) must be turned off and replaced by $E_0$. For the input fields $E_1$ and $E_2$, the symmetric phase pair, $\zeta$ and $\zeta'$, is provided by the product of the detuning $\pm\Delta$ and the pulse duration T/2: $\zeta = \Delta T/2$; $\zeta' = -\Delta T/2$. Figure 1c shows how to generate symmetric detuning $\pm\Delta$ using an acousto-optic modulator (AOM) driven by an rf-field generator. Figure 1d shows how both oppositely diffracted pulses are alternatively selected and combined with the original one, as seen in Fig. 1b. All controls are classical, deterministic, and compatible with current optoelectronic technologies.

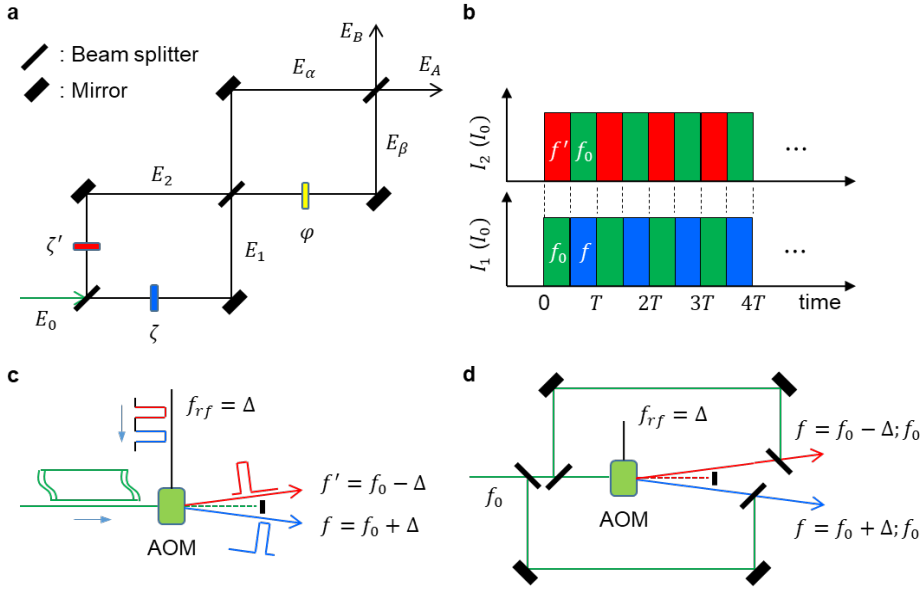

**Figure 1**| Schematic of macroscopic entangled field generation. **a.** A Mach-Zehnder interferometer for Hong-Ou-Mandel type proof. **b.** Alternative pulse sequence. **c.** Symmetric detuning. **d.** Superposition for basis randomness.

*Theory*

Based on Fig. 1, we now present a novel theory of macroscopically entangled light-pair generation. Using matrix representations for coherence optics, the following relations are obtained (see the Supplementary Information):

$$E_A = \frac{E_0}{2\sqrt{2}}\left[e^{i\zeta}(1-e^{i\varphi}) - e^{-i\zeta'}(1+e^{i\varphi})\right], \quad (1)$$

$$E_B = \frac{iE_0}{2\sqrt{2}}\left[e^{i\zeta}(1+e^{i\varphi}) - e^{-i\zeta'}(1-e^{i\varphi})\right]. \quad (2)$$

where $\zeta = \Delta T/2$, and $\zeta' = -\zeta$. The detuning $\Delta$ is with respect to the center frequency $f_0$ of the input field $E_0$, as shown in Fig. 1c, by an acousto-optic modulator (AOM) driven by an rf frequency at $f_{rf}$ ($=\Delta$), in which $f = f_0 + \Delta$ and $f' = f_0 - \Delta$. As a result, the corresponding intensities of the output fields are obtained:

$$I_A = \frac{I_0}{2}[1 - \sin(\varphi)\sin(\zeta;\zeta')], \quad (3)$$

$$I_B = \frac{I_0}{2}[1 + sin(\varphi)sin(\zeta;\zeta')], \quad (4)$$

where $sin(\zeta;\zeta')$ stands for either $sin(\zeta)$ or $sin(\zeta')$ via superposition with the original field as shown in Figs. 1c and 1d. The symmetric detuning control of $\pm\Delta$ by an AOM is for toggle switching between $f$ and $f'$ (see Fig. 1b). Thus, each mean value of the output intensity becomes uniform at $\langle I_A \rangle = I_0/2$ and $\langle I_B \rangle = I_0/2$ if $\zeta = (2n+1)\pi/2$ and $\zeta' = -\zeta$, i.e., $\Delta T = (2n+1)\pi/2$, where $T/2$ is the pulse duration of $E_1$ and $E_2$. Once again, the modulated and superposed fields, $E_1$ and $E_2$, are accompanied by $E_0$ for basis randomness, as shown in Figs. 1b and 1d.

Finally, the intensity product R of the output fields in Fig. 1a is as follows:

$$R = I_A I_B = [1 - sin^2(\varphi)sin^2(\zeta;\zeta')]. \quad (5)$$

In equation (5), $R = [1 - sin^2(\varphi)]$ is satisfied for the specific condition of symmetric phase control with $\zeta = (2n+1)\pi/2$ and $\zeta' = -\zeta$, where this result is deterministic and a single-shot measurement. Although the mean values of $I_A$ and $I_B$ are constant at $I_0/2$, the product R sinusoidally oscillates as a function of $\varphi$. This is the quintessence of the present theory for nonclassical features of anticorrelation in a macroscopic regime, resulting in:

$$g^{(2)}(\varphi) = \frac{\langle I_A I_B \rangle}{\langle I_A \rangle \langle I_B \rangle} = \frac{1}{2}\langle 1 - sin^2(\varphi) \rangle, \quad (6)$$

where conventional variable $\tau$ for coincidence measurements is now replaced by $\varphi$ for coherence measurements due to its higher sensitivity. Equation (6) is robust with respect to the laser bandwidth $\delta\omega$ ($c\delta\omega^{-1} \gg \lambda$) and thus shows a definite evidence of coherence-based quantum correlation in an interferometric scheme.

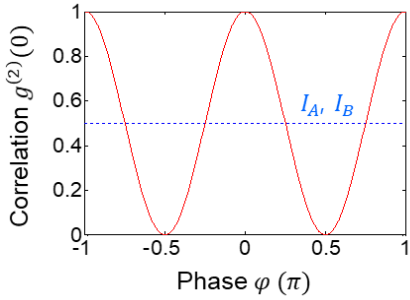

**Figure 2|** Numerical calculations for intensity correlation $g^{(2)}(\varphi)$. The phase $\varphi$ is within coincidence detection $\varphi = \zeta, \zeta'$.

Figure 2 shows numerical calculations for equation (6). As analyzed above, each output field's mean value is fixed at $I_0/2$ by an alternative selection of $\pm\frac{\pi}{2}$ phase-shifted $\zeta$ and $\zeta'$ using $\pm\Delta$ frequency control. Here, the intensity correlation $g^{(2)}(\varphi)$ covers both classical ($g^{(2)}(\varphi) \geq 0.5$) and quantum ($g^{(2)}(\varphi) < 0.5$) regimes depending on the $\varphi$ values. This is a unique feature of the wave property governed by the field's wavelength $\lambda$. Considering an actual bandwidth of coherent light $E_0$, however, equation (6) may result in dephasing for partial washout of the $g^{(1)}$ effect in $g^{(2)}$, as shown in Fig. 2b. Figure 1a is for monochromatic $E_0$ as a reference. Unlike the SPDC case with random initial phases among entangled photon pairs [29], the bandwidth-caused dephasing in equation (6) is far less sensitive. Thus, the present method of the macroscopically entangled light-pair generation is robust to laser sources. Experimental results will be presented elsewhere.

*Discussion*

In Fig. 1a, the specific condition for $\zeta = \pm\pi/2$ and $\zeta' = -\zeta$ is to compensate the BS-caused phase shift [41], resulting in uniform intensity distribution for $E_\alpha$ and $E_\beta$ via field bunching between $E_1$ and $E_2$:

$$E_1 = \frac{E_0}{\sqrt{2}}[1 - cos(\zeta;\zeta')], \quad (7)$$

$$E_2 = \frac{E_0}{\sqrt{2}}[1 + cos(\zeta;\zeta')]. \quad (8)$$

Due to the symmetric cosine function, equations (7) and (8) have no effect on the symmetric phase shift of $\pm\Delta$, resulting in $I_\alpha = I_\beta$ regardless of $\varphi$. Thus, $E_\alpha$ and $E_\beta$ impinging on a BS exhibit macroscopic quantum features, similar to a HOM dip with entangled photon pairs. For the proposed coherence anticorrelation in Fig. 1, the interference between $E_A$ and $E_B$ is $\varphi-$dependent, resulting in anticorrelation at $\varphi = \pm\frac{\pi}{2}(2n+1)$. Thus, the MZI in Fig. 1a acts as a quantum device whether the input field is a single photon or coherent light. As already discussed, anticorrelation in a HOM dip naturally satisfies the phase basis relation in a particular system [26,29]. For a BS, the phase bases for anticorrelation are $\pm\pi/2$, while for an MZI, it is 0 and $\pi$. In Fig. 1a, the phase basis is modified due to the $\zeta$ condition from $\pm n\pi$ to $\pm\pi/2$.

According to Heisenberg's uncertainty principle or de Broglie's wave-particle duality [42], conventional emphasis on the particle nature of photons is a matter of preference depending on the light source. Unlike SPDC-generated entangled photon pairs, the coherent light source has the benefits of determinacy and controllability. Due to such benefits of coherence optics, the confirmed entangled light pair $E_\alpha$ and $E_\beta$ can be extracted from the MZI system by inserting a BS into each arm, while keeping the same anticorrelation measurements for $E_A$ and $E_B$ in Fig. 1a. Compared with a typical laser system, this entangle light pair is called a quantum laser. Quantum mechanics is not as mysterious anymore in coupled systems, but instead can be definite and imperative as Einstein dreamed.

Regarding potential applications, the proposed method can be applied for a quantum laser whose light pair is macroscopically entangled, satisfying a N00N state with unbounded N. Compared with the MZI-superposition-based coherent de Broglie wavelength [27,30], the quantum laser has an additional benefit of robustness in phase fluctuations. The quantum laser may be applied for a quantum Lidar in quantum sensors, quantum keys in a quantum key distribution, and even a photonic qubit in quantum computations. Compared with amplitude-limited modulation in conventional quantum information, the proposed method may open the door to quantum phase modulation as well as quantum, wavelength division multiplexing. These applications are unprecedented and macroscopic in nature.

*Conclusion*

In conclusion, a novel method for macroscopically entangled light-pair generation was proposed, analyzed, and numerically demonstrated for both fundamental understanding of quantum mechanics and potential applications in future coherence-based quantum technologies. Unlike conventional understanding on quantum mechanics based on the particle nature of photons, the control of a coherent photon ensemble in the present analysis is phase deterministic in an MZI system for macroscopic quantum features. Owing to the wave nature of photons, coherence has also an inherent benefit of collective control, resulting in macroscopic quantum manipulation. The proposed method is compatible with coherence optics. The essential requirement for macroscopic quantum features is quantum superposition based on random phase bases, satisfying indistinguishability $g^{(1)}$ coherence as well as $g^{(2)}$ correlation. In other words, manipulation of macroscopic indistinguishability is a fundamental bedrock of quantum features that are achievable coherently. As defined in Bell's inequality, $g^{(1)}$ coherence has to be distinguished from classicality based on individual particles.

*Acknowledgments*
BSH acknowledges that the present work was supported by GIST via GRI-2021.